\documentclass[conference]{IEEEtran}
\IEEEoverridecommandlockouts

\usepackage[letterpaper, left=1in, right=1in, bottom=1in, top=0.75in]{geometry}
\usepackage{amsthm}
\usepackage{amsmath}
\usepackage{amssymb}
\usepackage{graphicx}
\usepackage{esint}

\makeatletter
\theoremstyle{plain}

\usepackage{amsfonts}
\usepackage{cite}
\usepackage{balance}
\usepackage{caption}
\usepackage{subcaption}
\usepackage{epstopdf}

\providecommand{\theoremname}{Theorem}

\makeatother

\providecommand{\theoremname}{Theorem}

\begin{document}

\author{\IEEEauthorblockN{Evangelos N. Papasotiriou\IEEEauthorrefmark{1}, Alexandros-Apostolos A. Boulogeorgos\IEEEauthorrefmark{1}, \\ Antigone Stratakou\IEEEauthorrefmark{1}, and  Angeliki Alexiou\IEEEauthorrefmark{1}
} \IEEEauthorblockA{\IEEEauthorrefmark{1}{\footnotesize{}{}{{{{{{{Department of Digital Systems, University of Piraeus, Piraeus 18534, Greece. }}}}}}}} 
\\
E-mails:  \{vangpapasot, alexiou\}@unipi.gr, antigonistratakou@hotmail.com, al.boulogeorgos@ieee.org 
} 
}
%
\title{Performance Evaluation of Reconfigurable Intelligent Surface Assisted\\ D--band Wireless Communication}
%
\IEEEoverridecommandlockouts
\IEEEpubid{\makebox[\columnwidth]{978-1-7281-7299-6/20/\$31.00 \copyright~2020~European~Union}
\hspace{\columnsep}\makebox[\columnwidth]{ }}
\maketitle


\begin{abstract}

In the recent years, the proliferation of wireless data traffic has led the scientific community to explore the use of higher unallocated frequency bands, such as the millimeter-wave (mmWave) and terahertz $\left(0.1-10\right)$ (THz) bands. However, they are prone to blockages from obstacles laid in the transceiver path. To address this, in this work the use of a reconfigurable-intelligent-surface (RIS) to restore the link between a transmitter (TX) and a receiver (RX), operating in the D--band $\left(110-170\text{ }\mathrm{ GHz}\right)$  is investigated. The system performance is evaluated in terms of pathgain and capacity considering the RIS design parameters, the TX/RX--RIS distance and the elevation angles from the center of the RIS to the transceivers.

\end{abstract}

\IEEEpeerreviewmaketitle

\section{Introduction}

In the past decade the wireless data traffic demand along with the need for high data rate transmissions has dramatically increased~\cite{Mag:Analytical_Performance_Assessment_of_THz_Wireless_Systems,PhD:Boulogeorgos,A:LC_CR_vs_SS,Mag:Bassar_Wir_Com_Through_RIS}. This has led the researchers to investigate the use of even higher frequency bands for beyond the fifth generation (5G) wireless technologies. Such frequency bands are D--band ($110$-$170\text{ }\mathrm{GHz}$) and Terahertz (THz) ($0.2$-$10\text{ }\mathrm{THz}$) band, because to date they can offer an abundance of unallocated bandwidth~\cite{ref1_VTC,Mag:High_Gain_D_Band_Transmitarrays,our_VTC_paper,WP:Wireless_Thz_system_architecture_for_networks_beyond_5G}.

The frequencies above $100\text{ }\mathrm{GHz}$ experience severe path attenuation, due to the free space propagation loss, as well as the electromagnetic (EM) wave energy absorption losses, by the molecules of the atmospheric medium. The molecular absorption loss of propagating EM waves at frequencies above $100\text{ }\mathrm{GHz}$ is mainly caused, by the atmospheric water-vapor molecules~\cite{Mag:Analytical_Performance_Assessment_of_THz_Wireless_Systems,C:ADistanceAndBWDependentAdaptiveModulationSchemeForTHzCommunications}. This makes the transmission above several cm without the of use highly directional antennas and line-of-sights (LoS) links impossible~\cite{Capacity_and_throughput_analysis_of_nanoscale_machine,our_VTC_paper,C:UserAssociationInUltraDenseTHzNetworks}. Furthermore, the small wavelength of higher frequencies, like those of the mmWave, sub-THz and THz bands, reduces their ability to diffract around obstacles with size several times their wavelength~\cite{C:Relay-Based_Blockage_and_Antenna_Misalignment_Mitigation_in_THz_Wireless_Communications},~\cite{C:mmWave_Human_Blockage_at_73_GHz_with_a_Simple_Double_Knife_Edge_Diffraction_Model_and_Extension_for_Directional_Antennas}, which can lead to LoS link blockage. 

In order to address this limitation and re-establish the link between the transmitter (TX) and the receiver (RX), various technical papers have proposed the use of reconfigurable-intelligent-surfaces (RIS)~\cite{Let:Int_RISs:Phys_Prop_PL_Mod,Arx:Wir_Com_with_RIS_PL_Mod,Arx:Mod_and_Anal_of_RIS_for_Ind_and_Out_Appl_in_6G_Wir_Sys,
Trans:LIS:Assisted_Wir_Com_Expl_Stat_CSI,Trans:Perf_Anal_of_LIS:Asym_Data_Rate_and_Ch_Hard_Ef,Mag:Bassar_Wir_Com_Through_RIS,
C:RIS_vs_Rel_Diffs_Sim_and_Per_Comp,Mag:Perf_Anal_of_RIS_Assist_Wir_Sys_and_Comp_with_Relaying}. A RIS is an artificial surface capable of steering the propagation of the impinging signal to a desired direction~\cite{Mag:Bassar_Wir_Com_Through_RIS},~\cite{Arx:Wir_Com_with_RIS_PL_Mod},~\cite{Mag:Smart_radio_env_emp_by_RIS_AI}. Furthermore, a RIS can be engineered as a very thin sheet of EM material with size smaller than the wavelength~\cite{Mag:Bassar_Wir_Com_Through_RIS},~\cite{Arx:Wir_Com_with_RIS_PL_Mod},~\cite{Synth_of_EM_Metasurfaces_Principles_and_Illustrations} and is manufactured to possess properties that cannot occur in natural materials. Specifically, a RIS can manipulate the amplitude and phase of the impinging wave, by having the ability to change them in a different manner than the natural occurring specular reflection. These materials are called meta-materials and the resulting RIS is known as a meta-surface~\cite{Mag:Bassar_Wir_Com_Through_RIS},~\cite{Let:Int_RISs:Phys_Prop_PL_Mod}. In more detail, the authors in~\cite{Arx:Wir_Com_with_RIS_PL_Mod} have modeled the free-space-path-loss (FSPL) of RIS assisted wireless links under different scenarios, by considering the physics and EM nature of the RISs. Furthermore the RIS was modeled as a two-dimensional (2--D) antenna array of meta-surface elements. The proposed FSPL models enclosed the RIS dimensions, the near-field/far-field effects of the RIS--transceiver separation distance and the radiation patterns of transceiver antennas and RIS unit cells. In~\cite{Let:Int_RISs:Phys_Prop_PL_Mod}, the far-field FSPL of a 2--D RIS assisted wireless communication scenario is modeled, by using physical optics techniques. More specifically, it is first explained how, a passive metallic surface scatters the incident wave and then, how a RIS should be designed in order to mimic such a surface, while also controlling the directivity of the scattered wave. In~\cite{Mag:Bassar_Wir_Com_Through_RIS}, the authors presented a single-input-single-output (SISO) wireless communication scheme through a single RIS constructed as an array of meta-surface elements. In more detail, two RIS usage scenarios were presented, where it was assumed that the RIS consisted of a large number of elements and only through the RIS the signal was conveyed from source to destination. In the first scenario the RIS was assumed to be able to establish LoS links with both TX and RX. In the second, not only it could establish LoS links with the transceivers, but also provided that the RIS is sufficiently close to an RF wave generator it can be used as a TX, by modulating the impinging signal. Then in both cases the signal-to-noise-ratio (SNR) distribution of the RX was derived and the underlying system performance was evaluated in terms of symbol-error-probability (SEP). As an extension of~\cite{Mag:Bassar_Wir_Com_Through_RIS}, in~\cite{Arx:Mod_and_Anal_of_RIS_for_Ind_and_Out_Appl_in_6G_Wir_Sys}, 
the use of multiple RISs as a means to establish TX--RX links in non-LoS (NLoS) conditions in sub ${6\text{ GHz}}$ to mmwave frequencies is motivated. Then, the developed transmission models are used with empirical pathloss models and the performance is evaluated in terms of SEP, asymptotic data rate and capacity. Moreover, in~\cite{Trans:Perf_Anal_of_LIS:Asym_Data_Rate_and_Ch_Hard_Ef}, an uplink multi-user scenario from the user equipments (UEs) to the RIS was investigated. The received signal at the RIS was assumed to experience estimation errors, interference due to spatially correlated fading channels and RIS hardware impairments. The system performance was evaluated in terms of capacity and an upper bound on ergodic capacity (EC). Also, in~\cite{Trans:LIS:Assisted_Wir_Com_Expl_Stat_CSI}, a multiple-input-single-output (MISO) RIS aided wireless communication scenario was presented, where the RIS was modeled as a uniform linear array of metasurface elements. Then, an optimal phase shift design was proposed based on the upper bound of the EC and statistical channel state information. Furthermore, in~\cite{C:RIS_vs_Rel_Diffs_Sim_and_Per_Comp}, the use of RIS vs decode (DF) and amplify (AF) and forward relays was investigated for mmwave and sub-mmwave links. The performance of the respective systems was evaluated in terms of spectral efficiency. Additionally, in~\cite{Mag:Perf_Anal_of_RIS_Assist_Wir_Sys_and_Comp_with_Relaying}, the authors provided the theoretical framework for the performance comparison of RISs and AF relaying wireless systems. In more detail, the closed form expressions for the instantaneous and average end-to-end SNR of both systems was extracted. Then, building upon these expressions the performance of both investigated systems was evaluated in terms of SEP, outage probability (OP) and EC.

Motivated by the above, in this paper, we investigate the performance of a D--Band ($100-170$) GHz downlink (DL) wireless link, which has the direct LoS link between the TX and RX blocked by obstacles. Specifically, the signal is conveyed through a 2--D RIS array of metasurface elements, that is assumed to establish LoS links with both TX and RX. Moreover, the system performance is evaluated in terms of pathgain (PG) and capacity. In more detail, these metrics are given as a function of the number of RIS unit cell elements, the unit cell size, the TX--RIS and RX--RIS center distances, the transceiver antenna gains, and elevation angles from the RIS center to the TX and RX antennas, respectively. The aforementioned parameters and performance metrics are investigated under the assumptions that the RIS is at the far-field of the transceivers and the metasurface operates as a specular reflector.

\section{System \& Channel Model}\label{S:SM}

In this work a DL D--Band indoor scenario is considered, where the LoS link between the TX and RX is assumed to be interrupted by various obstacles (such as people, furnitures, plaster boards, etc). The wireless transmission is achieved by using a  single RIS able to establish LoS links with both the TX and RX. Also, both the TX and RX are equipped with a single directive antenna, which is assumed to able to create one major lobe and several minor lobes of insignificant beamwidth.

\begin{figure}
\centering\includegraphics[width=0.99\linewidth,trim=0 0 0 0,clip=false]{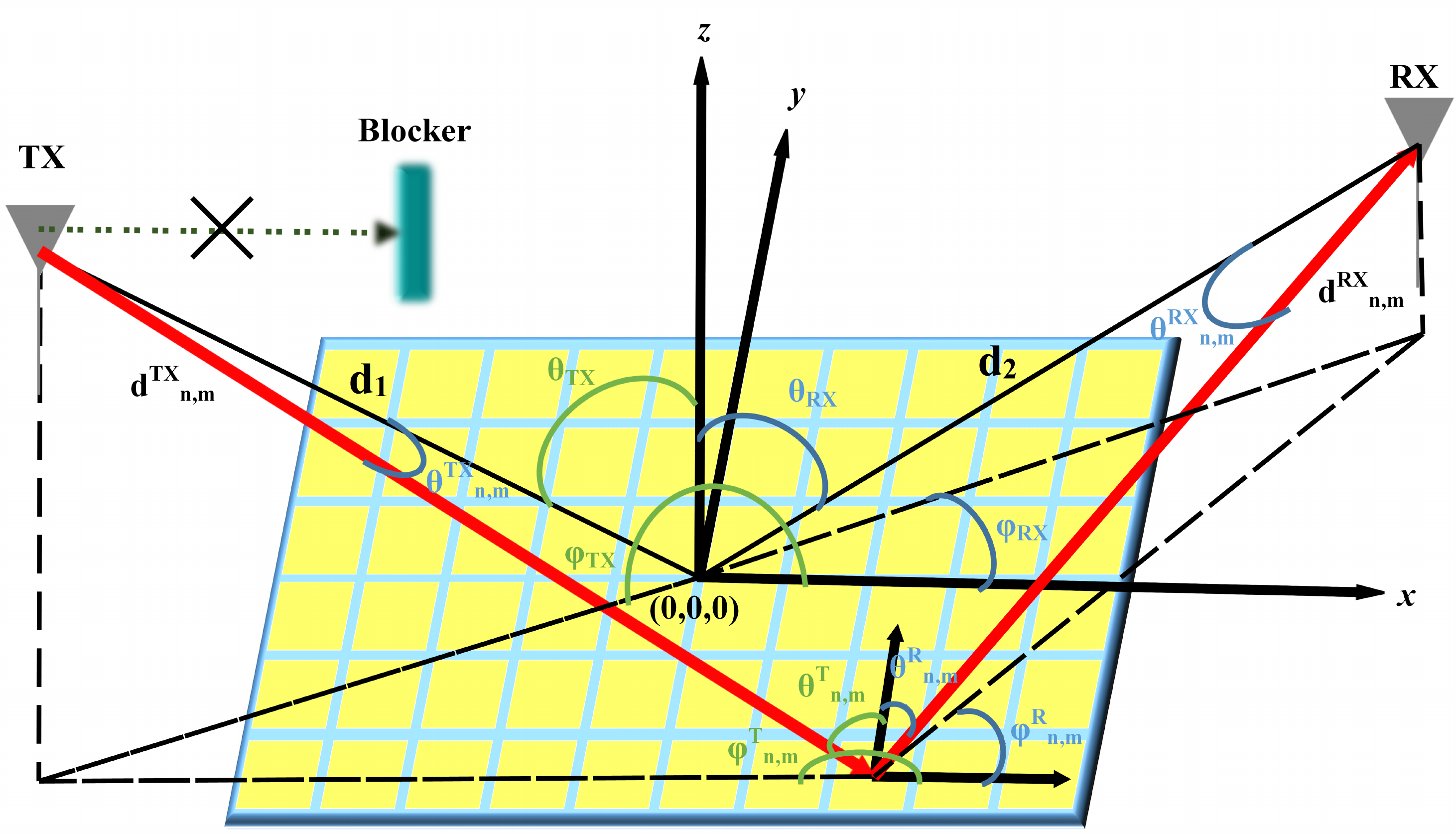}
\caption{RIS-assisted D--band wireless communication with the TX--RX LoS path blocked.}
\label{fig:System_model}
\end{figure}
As shown, in Fig.~\ref{fig:System_model}, the RIS is placed on the ${x-y}$ plane of a three dimensional (3D) Cartesian coordinate system centered at $(0,0,0)$, while the TX and RX are located at $(x_{TX},y_{TX},z_{TX})$ and $(x_{RX},y_{RX},z_{RX})$, respectively. Furthermore, the RIS consists of ${N\times M}$ unit cell elements placed in an orthogonal grid with $N$ rows and $M$ columns. Each unit cell is denoted as $K_{n,m}$, has dimensions $d_x$ and $d_y$ along the $x$ and $y$ axis, respectively and is centered at $\left(\left(m-\frac{1}{2}\right)d_x,\left(n-\frac{1}{2}\right)d_y,0\right)$, where $n\in\left[1-\frac{N}{2},\frac{N}{2}\right]$, $m\in\left[1-\frac{M}{2},\frac{M}{2}\right]$. It should be noted that $d_x$ and $d_y$ are in sub-wavelength order and are commonly designed in the range $\left[\frac{\lambda}{10},\frac{\lambda}{2}\right]$~\cite{Mag:Bassar_Wir_Com_Through_RIS},~\cite{Arx:Wir_Com_with_RIS_PL_Mod} and~\cite{Synth_of_EM_Metasurfaces_Principles_and_Illustrations} (and references therein), and $\lambda$ stands for the wavelength. Furthermore, the distances of the RIS and $K_{n,m}$ centers with the TX and RX equal $d_1$, $d_2$, $d^{TX}_{n,m}$ and $d^{RX}_{n,m}$, respectively. Additionally, the elevation and azimuth angles from the RIS and $K_{n,m}$ centers to the TX and RX equal $\theta_{TX}$, $\phi_{TX}$, $\theta_{RX}$, $\phi_{RX}$, $\theta^T_{n,m}$, $\phi^T_{n,m}$, $\theta^R_{n,m}$ and $\phi^R_{n,m}$, respectively. Finally, the elevation and azimuth angles from the TX and RX to $K_{n,m}$ are denoted as $\theta^{TX}_{n,m}$, $\phi^{TX}_{n,m}$, $\theta^{RX}_{n,m}$ and $\phi^{RX}_{n,m}$, respectively.

The baseband equivalent received signal can be expressed~as
\begin{align}
y=hx+n,
\label{Eq:rec_signal}
\end{align}
where $x$, $h$ and $n$ stand for the transmitted signal, the channel coefficient and the additive white Gaussian noise (AWGN) with variance $N_o$, respectively. In more detail, the channel gain $|h|$ can be obtained as~\cite{Arx:Wir_Com_with_RIS_PL_Mod},~\cite{Mag:Con_anal_of_RIS_assist_THz_wir_sys}
\small
\begin{align}
\begin{split}
&|h|=\sqrt{\frac{G_t G_r G d_x d_y}{64 \pi^3}}\lambda
\left|
\sum _{m=1-\frac{M}{2}}^{\frac{M}{2}} \sum _{n=1-\frac{N}{2}}^{\frac{N}{2}}
\frac{\sqrt{F_c}\Lambda_{n,m}}{d^{TX}_{n,m}d^{RX}_{n,m}} \right. \times \\
& \left. \exp\left(-j\frac{2\pi\left(d^{TX}_{n,m}+d^{RX}_{n,m}\right)}{\lambda}\right) \sqrt{\tau\left(f,d^{TX}_{n,m}\right)\tau\left(f,d^{RX}_{n,m}\right)}
\right|,
\end{split}
\label{Eq:h_gain_gen_formula}
\end{align}
\normalsize
\begin{align}
\Lambda_{n,m}=A_{n,m}\exp\left(j\phi_{n,m}\right),
\label{Eq:el_coefficient}
\end{align}
\begin{align}
d^{TX}_{n,m}=\sqrt{\gamma+\delta+z^2_{TX}},
\label{Eq:d_TX_n_m}
\end{align}
\begin{align} 
d^{RX}_{n,m}=\sqrt{\alpha+\beta+z^2_{RX}},
\label{Eq:d_RX_n_m}
\end{align}
\begin{align}
\gamma=\left(x_{TX}-\left(m-\frac{1}{2}\right)d_x\right)^2,
\end{align}
\begin{align}
\delta=\left(y_{TX}-\left(n-\frac{1}{2}\right)d_y\right)^2,
\end{align}
\begin{align}
\alpha=\left(x_{RX}-\left(m-\frac{1}{2}\right)d_x\right)^2,
\end{align}
\begin{align}
\beta=\left(y_{RX}-\left(n-\frac{1}{2}\right)d_y\right)^2,
\end{align}
\begin{align}
\begin{split}
F_c &=F^{TX}(\theta^{TX}_{n,m},\phi^{TX}_{n,m}) F(\theta^T_{n,m},\phi^T_{n,m}) \times \\ 
& F(\theta^R_{n,m},\phi^R_{n,m})F^{RX}(\theta^{RX}_{n,m},\phi^{RX}_{n,m}),
\label{Eq:Fc}
\end{split}
\end{align}
\begin{align}
F\left(\theta,\phi\right)=\begin{cases}
 cos^x\left(\theta\right) & \theta\in\left[0,\text{ }\frac{\pi}{2}\right], \phi\in\left[0,2\pi\right] \\
 0 & \theta\in\left(\frac{\pi}{2},\pi\right],\text{ } \phi\in\left[0,2\pi\right]
\end{cases},
\label{Eq:Normalized_power_radiation_pattern}
\end{align}
\begin{align}
G_{t,r}=\frac{4 \pi}{\int _{0}^{2 \pi}\int _{0}^{\frac{\pi}{2}}F\left(\theta,\phi\right)sin\theta d\theta d\phi},
\label{Eq:Antenna_Gain}
\end{align}
where $f$ stands for the frequency, $\Lambda_{n,m}$ denotes the programmable reflection coefficient of $K_{n,m}$, while $A_{n,m}$ and $\phi_{n,m}$ are the amplitude and phase, respectively. Furthermore, the parameters $G_t$, $G_r$ and $G$ stand for the TX, RX antenna gains and $K_{n,m}$ gain, respectively and are obtained from~\eqref{Eq:Antenna_Gain}, as shown in~\cite{Mag:Con_anal_of_RIS_assist_THz_wir_sys}~and~\cite{B:Ant_Th_and_Des}. Moreover, $\tau(f,d)$ represents the transmittance of the absorbing atmospheric medium~\cite{Mag:Analytical_Performance_Assessment_of_THz_Wireless_Systems},~\cite{C:100_450GHz_ch_model_joonas_2020},~\cite{Mag:Perf_Anal_of_THz_Wir_Sys_in_the_Pres_of_Ant_Misal_and_PHN}, which shows the fraction of the energy of the propagating EM wave capable of reaching the RX antenna without being absorbed by the intermediate molecules. Furthermore, $F_c$ stands for the product of the normalized power radiation patterns, $F(\cdot,\cdot)$ obtained from~\eqref{Eq:Normalized_power_radiation_pattern}, as also shown in~\cite{Arx:Wir_Com_with_RIS_PL_Mod},~\cite{Mag:Con_anal_of_RIS_assist_THz_wir_sys} and~\cite{B:Ant_Th_and_Des}. Moreover, by assuming that $d_1$ and $d_2$ are in the far-field of the RIS~\cite{Arx:Wir_Com_with_RIS_PL_Mod},~\cite{B:Antennas_from_theory_to_practice}, the direction of peak radiation of both the TX and RX antennas point at the center of the RIS, i.e. ${F^{TX}\left(\theta^{TX}_{n,m},\phi^{TX}_{n,m}\right)\approx1}$ and ${F^{RX}\left(\theta^{RX}_{n,m},\phi^{RX}_{n,m}\right)\approx1}$. Additionally, all $K_{n,m}$ unit cells share the same reflection coefficient $\Lambda_{n,m}=A\exp\left(j\phi_{n,m}\right)$ and by employing~\eqref{Eq:d_TX_n_m}~and~\eqref{Eq:d_RX_n_m},~\eqref{Eq:h_gain_gen_formula} can expressed~as
\begin{align}
\begin{split}
|h|& = \sqrt{\frac{B F\left(\theta_{TX},\phi_{TX}\right) F\left(\theta_{RX},\phi_{RX}\right)}{64 \pi^3 d_1^2 d_2^2}}
\left| \frac{\Phi}{X} \frac{\Psi}{\Omega}\right|,
\end{split}
\label{Eq:h_gain_far_field_formula}
\end{align}
where
\begin{align}
B\left(f\right)=G_t G_r G M^2 N^2 d_x d_y \lambda^2 A^2 \tau\left(f,d_1+d_2\right),
\label{Eq:Beta_f}
\end{align}
\begin{align}
\begin{split}
\Phi & = sinc\left(\frac{M\pi}{\lambda}\left(sin\left(\theta_{TX}\right)cos\left(\phi_{TX}\right) \right. \right. \\
& \left. \left. + sin\left(\theta_{RX}\right)cos\left(\phi_{RX}\right)\right)d_x \vphantom{\frac{M\pi}{\lambda}}\right),
\end{split}
\label{Eq:Phi}
\end{align}
\begin{align}
\begin{split}
X&=sinc\left(\frac{\pi}{\lambda}\left(sin\left(\theta_{TX}\right)cos\left(\phi_{TX}\right) \right. \right. \\
& \left. \left. +sin\left(\theta_{RX}\right)cos\left(\phi_{RX}\right)\right)d_x \vphantom{\frac{\pi}{\lambda}} \right),
\end{split}
\label{Eq:X}
\end{align}
\begin{align}
\begin{split}
Y&=sinc\left(\frac{N\pi}{\lambda}\left(sin\left(\theta_{TX}\right)sin\left(\phi_{TX}\right) \right. \right. \\
& \left. \left. +sin\left(\theta_{RX}\right)sin\left(\phi_{RX}\right)\right)d_y \vphantom{\frac{N\pi}{\lambda}}\right),
\end{split}
\label{Eq:Y}
\end{align}
\begin{align}
\begin{split}
\Omega&=sinc\left(\frac{\pi}{\lambda}\left(sin\left(\theta_{TX}\right)sin\left(\phi_{TX}\right) \right. \right. \\
& \left. \left. +sin\left(\theta_{RX}\right)sin\left(\phi_{RX}\right)\right)d_y \vphantom{\frac{\pi}{\lambda}} \right).
\end{split}
\label{Eq:Omega}
\end{align}
Then, from~\eqref{Eq:h_gain_far_field_formula} by assuming that the RIS is manipulated as a specular reflector, i.e. ${\theta_{RX}=\theta_{TX}}$ and $\phi_{RX}=\phi_{TX}+\pi$, $|h|$ can be expressed~as
\begin{align}
|h|=\sqrt{\frac{B F\left(\theta_{TX},\phi_{TX}\right)F\left(\theta_{RX},\phi_{RX}\right)}{64 \pi^3 d_1^2 d_2^2}}.
\label{Eq:h_gain_spec_reflection}
\end{align}
\section{Performance Evaluation}\label{S:PE}

In order to quantify the potential of using a RIS to restore the LoS link between a TX and RX operating in the D--band, which is blocked by obstacles, the fundamental link quality performance metric of capacity is presented. In more detail, to evaluate the wideband channel capacity in the D--band, the received signal is decomposed in $W$ narrow sub--bands of bandwidth $\Delta f_i$, where $i\in\left[1,W\right]$. The sub--bands are employed in order to ensure a flat fading wireless channel~\cite{jornet2011},~\cite{Mag:Multi-Ray_Ch_Mod_and_Wid_Char_for_Wir_Com_THz}. Therefore, the capacity can be obtained~as
\begin{align}
C=\sum _{i=1}^W \Delta f_i log_2\left(1+SNR_i\right),
\label{Eq:capacity}
\end{align}
\begin{align}
SNR_i=\frac{|h_i|^2 P_{t_{i}}}{N_o},
\label{Eq:SNR_i}
\end{align}
where $SNR_i$, $h_i$ and $P_{t_i}$ stand for the SNR, the channel coefficient and power spectral density (PSD) of the transmitted signal in the $i$--th sub--band, under the constraint $\sum _{i=1}^{W} P_{t_i} \leq P$, with $P$ representing the total transmitted power by the TX. Furthermore, using~\eqref{Eq:h_gain_spec_reflection} in~\eqref{Eq:SNR_i},~\eqref{Eq:capacity} can be equivalently rewritten~as
\begin{align}
C = \sum _{i=1}^W \Delta f_i log_2\left(1+\Xi_i \frac{P_{t_{i}}}{N_o}\right),
\label{Eq:capacity_spec_refl}
\end{align}
where $\Xi_i=\frac{B\left(f_i\right) F\left(\theta_{TX},\phi_{TX}\right) F\left(\theta_{RX},\phi_{RX}\right)}{64 \pi^3 d_1^2 d_2^2}$.

\section{Results \& Discussion}\label{S:RD}

In this section the effectiveness of a D--band RIS assisted wireless link is evaluated in terms of PG, transceiver normalized power radiation pattern and capacity. The D--band LoS link between the TX and RX is assumed to be fully blocked by obstacles and the wireless transmission is conveyed over a single RIS capable of establishing LoS links with both the TX and RX. Also, in the following results the RIS is assumed to operate as a specular reflector with ${\phi_{n,m}=0}$ and ${A_{n,m}=A=1}$. It should be noted, that setting $A=1$ is a practical assumption, because the RIS unit cells are designed to maximize the signal reflection~\cite{Let:IRS_Enh_OFDM_Ch_Est_and_Refl_Opt},~\cite{Let:Spec_and_En_Ef_of_IRS_As_MISO_Com_with_HD_Imp}. Furthermore, it is assumed that ${N=M}$, ${d_1=d_2}$, ${\theta_{RX}=\theta_r}$, ${\theta_{TX}=\theta_t}$, ${\phi_{TX}=\phi_t}$, ${\phi_{RX}=\phi_r}$, ${\theta_r=\theta_t}$ and ${\phi_r=\phi_t+\pi}$. Additionally, ${G_t=G_r\in\left[20,37\right]}$ dB~\cite{Mag:High_Gain_D_Band_Transmitarrays},~\cite{Mag:D_Band_H_Gain_Circ_Pol_Plate_Array_Ant},~\cite{C:Two_Types_of_High_Gain_Slot_Array_Ant_based_on_Ridge_Gap_Waveguide_D_Band} (these are some indicative transceiver gains measured in the D--band and the corresponding antenna designs are not taken into account in this work) and $G=10$ dB. Moreover, the maximum RIS dimensions are assumed to be ${10 \lambda \times 10 \lambda}$. Also, standard atmospheric conditions are assumed, i.e. temperature $T=296$ Kelvin, relative humidity $\chi=50 \%$ and atmospheric pressure $p=101325$ Pa. Finally, $P$ is assumed to be equally divided and allocated to each $\Delta f_i$. 

\begin{figure}
\centering\includegraphics[width=0.6\linewidth,trim=0 0 0 0,clip=false]{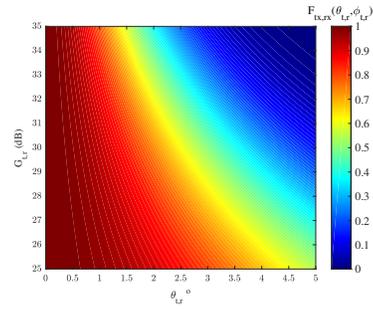}
\caption{TX/RX normalized power radiation pattern as a function of $\theta_t$, $\theta_r$ and $G_t$, $G_r$.}
\label{Fig:Normalized_PRP}
\end{figure}

Fig.~\ref{Fig:Normalized_PRP}, demonstrates the TX/RX normalized power radiation patterns as a function of $\theta_{t,r}$ and $G_{t,r}$, where due to the specular reflection assumption ${F\left(\theta_t,\phi_t\right)=F\left(\theta_r,\phi_r\right)}$. In more detail, it is observed that for a given $G_{t,r}$ as $\theta_{t,r}$ increases the normalized power radiation pattern significantly decreases. For example, for $G_{t,r}=30$ dB and changing $\theta_{t,r}$ from $1^o$ to $4.5^o$ the normalized power radiation pattern decreases from $0.9$ and $0.21$, respectively. Also, for a given $\theta_{t,r}$ as $G_{t,r}$ increases the normalized power radiation pattern significantly decreases. As an example, for $\theta_{t,r}=2.5^o$
and changing $G_{t,r}$ from $25$ to $35$ dB the normalized power radiation pattern decreases from $0.87$ to $0.22$, respectively. Finally, since $F\left(\theta_t,\phi_t\right) \times F\left(\theta_r,\phi_r\right)$ is proportional to the PG it is observed that $\theta_{t,r} \leq 1.6^o$ is needed in order to achieve $F\left(\theta_{t,r},\phi_{t,r}\right)\geq0.6$.

\begin{figure}
\centering\includegraphics[width=0.6\linewidth,trim=0 0 0 0,clip=false]{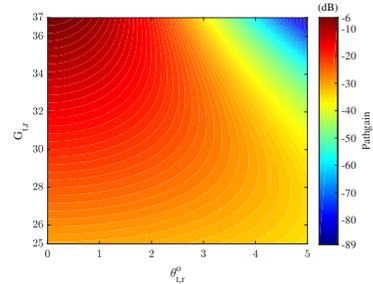}
\caption{Pathgain as a function of $\theta_t$, $\theta_r$ and $G_t$, $G_r$.}
\label{Fig:PG_vs_tht_vs_Gt}
\end{figure}

Fig.~\ref{Fig:PG_vs_tht_vs_Gt} illustrates the PG as function of the TX/RX elevation angles and antenna gains, assuming $d_1=d_2=2.5$ m, $f=110\text{ GHz}$, ${d_x=d_y=\frac{\lambda_{110}}{10}=0.00027\text{ m}}$ and ${M^2=12.1\times10^3}$. It is observed that for a given $G_{t,r}$ as $\theta_{t,r}$ increases the PG decreases. Specifically, it is shown that even though by increasing $G_{t,r}$, the PG does not always improve, on the other hand it deteriorates with the increase of $\theta_{t,r}$. 
As an example for $G_{t,r}=25$ dB and by increasing $\theta_{t,r}$ from $1^o$ to $3^o$ and $5^o$ the PG changes from $-30.03$ to $-9.14$ and $-31.71$ dB, respectively, whereas for $G_{t,r}=37$ dB and the same values of $\theta_{t,r}$ the PG is $-35.86$, $-35$ and $-88.78$ dB, respectively. This is because for $G_{t,r}=25$ dB and $\theta_{t,r}$ equal to $1^o$, $3^o$ and $5^o$, $F\left(\theta_{t,r},\phi_{t,r}\right)$ changes from $0.97$ to $0.79$ and $0.8$, respectively, whereas for $G_{t,r}=37$ dB and the same values of $\theta_{t,r}$, $F\left(\theta_{t,r},\phi_{t,r}\right)$ equals $0.11$, $0.55$ and $0.002$, respectively.

\begin{figure}
\centering\includegraphics[width=0.6\linewidth,trim=0 0 0 0,clip=false]{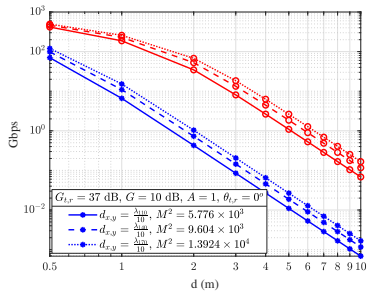}
\caption{Capacity as a function of $d$ for different values of $d_{x,y}$ and $\frac{P}{N_o}$.}
\label{Fig:Capacity_vs_d_diff_dx}
\end{figure}

In Fig.~\ref{Fig:Capacity_vs_d_diff_dx}, the capacity is given as function of $d$ for different values of $d_{x,y}$ and $\frac{P}{N_o}$. In more detail, the blue and red lines stand for ${\frac{P}{N_o}=5}$ and ${\frac{P}{N_o}=25}$ dB, respectively, while the continuous, dashed and dotted lines represent $d_{x,y}$ equal to $0.00027$, $0.00021$ and $0.0002$ m, respectively. It is observed that as $d$ increases the capacity decreases, because $d_1^2 d_2^2$ is inversely proportional to the PG. Moreover, for a given $d$ and $d_{x,y}$ by increasing $\frac{P}{N_o}$ the capacity increases. For example, for ${d=2}$ m, ${d_{x,y}=0.0002}$ m and by changing $\frac{P}{N_o}$ from $5$ to $25$ dB, the capacity increases from $1.038$ to $67.71$ Gbps, respectively. Moreover, for a given $d$ and $\frac{P}{N_o}$, as $d_{x,y}$ decreases the capacity increases. This is due to the fact that the smaller the $d_{x,y}$, the more the elements that can be packed in the RIS. As a result, greater $M^2$ improves the PG and capacity. As an example, for 
$d=1$ m and $\frac{P}{N_o}=25$ dB by changing $d_{x,y}$ from $0.00027$ to $0.0002$ m, results in ${M^2=5.776 \times 10^3}$ and ${M^2=1.3924 \times 10^4}$, respectively, achieving a capacity of $187.5$ and $258.2$ Gbps, respectively.

\begin{figure}
\centering
\includegraphics[width=0.6\linewidth,trim=0 0 0 0,clip=false]{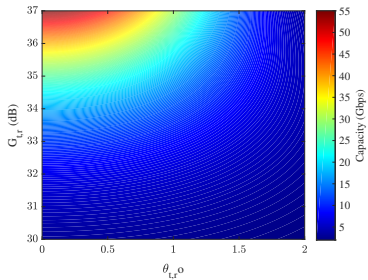}
\caption{Capacity as function of $\theta_t$, $\theta_r$, $G_t$ and $G_r$.}
\label{Fig:Capacity_vs_tht_vs_Gt}
\end{figure}

In Fig.~\ref{Fig:Capacity_vs_tht_vs_Gt}, the capacity is given as a function of $\theta_{t,r}$ and $G_{t,r}$ for ${d=2.5}$ m, ${\frac{P}{N_o}=25}$ dB, ${d_{x,y}=\frac{\lambda_{110}}{10}}=0.00027$ m and ${M^2=12.1 \times 10^3}$. It is observed that for given $G_{t,r}$ as $\theta_{t,r}$ increases the capacity significantly decreases. For example, for $G_{t,r}=37$ dB and  by changing $\theta_{t,r}$ from $0^o$ to $0.5^o$, $1^o$ and $2^o$, the capacity is $54.5$, $46.15$, $29.85$ and $3.5$ Gbps, respectively. This is because $F\left(\theta_{t,r},\phi_{t,r}\right)\approx0$ with the increase of $\theta_{t,r}$. Furthermore, for small values of $\theta_{t,r}$ as $G_{t,r}$ increases the capacity improves. However, for high values of $\theta_{t,r}$ increasing $G_{t,r}$ degrades the capacity. Moreover, for example for ${\theta_{t,r}=0^o}$ and increasing $G_{t,r}$ from $30$ to $32.3$, $34.25$, $35.65$ and $37$ dB, yields a capacity of $3.015$, $8.41$, $19.29$, $33.57$, $54.5$, while for ${\theta_{t,r}=0.6^o}$ and the same $G_{t,r}$ values the capacity increases from $2.85$ to $7.69$, $16.92$, $28.38$ and $44.32$ Gbps, respectively. Additionally, for ${\theta_{t,r}=1^o}$ and the same $G_{t,r}$ increase, the achieved capacity is $2.59$, $6.57$, $13.35$, $20.81$, $29.85$ Gbps respectively, whereas for ${\theta_{t,r}=2^o}$ and the aforementioned values of $G_{t,r}$ the capacity changes from $1.65$ to $3.09$, $4.19$, $4.3$ and $3.56$ Gbps, respectively. Finally, in order to achieve capacity greater than $18$ Gbps, it is observed that ${G_{t,r} \geq 35}$ dB and ${\theta_{t,r} \leq 1.2^o}$ are required, i.e. $F\left(\theta_{t,r},\phi_{t,r}\right) \geq 0.7$.

\section{Conclusions}\label{S:Conclusions}

In this work, the performance of a RIS assisted D--band wireless link was presented in terms of pathgain and capacity. In more detail, the metrics were presented as a function of frequency, the RIS number of unit cells, the TX/RX--RIS center distance, the RIS element size, the transceiver antenna gains and the elevation angles from the TX/RX to the RIS center. Furthermore, the results revealed that for a given $G_{t,r}$ as $\theta_{t,r}$ increases the normalized power radiation patterns of the TX and RX decrease. This indicates that the transceiver gain increase improves the system performance, when $\theta_{t,r}$ is low. Also, smaller $d_{x,y}$ yield more unit cell elements packed within the RIS, which acts as a gain and improves the system performance. Additionally, as expected decreasing the TX--RIS, RIS--RX distances and increasing $\frac{P}{N_o}$ yields better pathgain and capacity. Finally, in order to achieve capacity greater than $18$ Gbps, ${G_{t,r} \geq 35}$ dB and ${\theta_{t,r} \leq 1.2^o}$ are required, i.e. $F\left(\theta_{t,r},\phi_{t,r}\right) \geq 0.7$. 

\section*{Acknowledgement}

This work has received funding from the European Commission Horizon 2020 research and innovation programme ARIADNE under grant agreement No. 871464.

\balance
\bibliographystyle{IEEEtran}
\bibliography{References}

\end{document}